# Mean Square Performance of a family of Adaptive Algorithms for colored noise

R. Sankara Prasad
*Department of Electrical & Electronics Engineering*
Birla Institute of Technology and Science
Pilani, Rajasthan, India
f2015257@pilani.bits-pilani.ac.in

*Abstract*— In real-time applications the characteristics and properties of a signal vary inconsistently. So, to maintain the integrity of such signals there is a need for effective adaptive filters. The conventional Least Mean Squared(LMS) algorithm is widely used because of its computational simplicity and ease of implementation. But, its convergence speed rapidly reduces when colored noise is present in the signal. Affine projection(AP) algorithms are used to speed up the convergence but have high computational costs. In this paper, the mean square performance of LMS and AP algorithms is analyzed when subject to white noise and colored noise.

## I. INTRODUCTION

An adaptive filter is a linear filter, whose coefficients can be modified using an optimization algorithm. Adaptive filters are required because, the parameters of the system we need to model many times are unknown or they keep changing. Adaptive filters are normally digital filters because of their computational complexity. It is a closed loop system where the error is sent as feedback to the optimization algorithm and is further used to update the coefficients of the filter.

Least Mean Square(LMS) algorithm is widely used as the preferred optimization algorithm because of its low cost of implementation while providing decent convergence rates for a variety of applications. But, the convergence rate of LMS algorithm reduces rapidly for colored inputs. Also, the convergence rate depends heavily on the amplitude of the input signal [1]. The basic affine projection algorithm proposed by Ozeki and Umeda attempts to increase convergence rates by projecting the input vector to the affine subspace [2]. Since then, many improvements and modifications have been proposed for a variety of applications like the Regularised Affine Projection(R-AP)[3] algorithm or the Binormalized Data Reusing LMS (BNDR-LMS)[4]. All these algorithms come under the AP algorithms family.The main drawback of the AP family is that the computational cost and hardware implementation are costly and difficults. But with the advent of fast DSP processors such algorithms are becoming a viable alternative.

In this paper, each of these algorithms are implemented for the application of system identification. First the response of the system is analysed for input white noise and then for colored noise. The different algorithms are compared on the basis of fast convergence, low mean square error and lesser computational complexity.

Throughout the paper, the following notations are adopted:
( )$^T$      Transpose of a vector or a matrix.
$\| \ \|$      Euclidean norm of a vector.
( )$^H$      Hermitian conjugation

## II. SYSTEM IDENTIFICATION

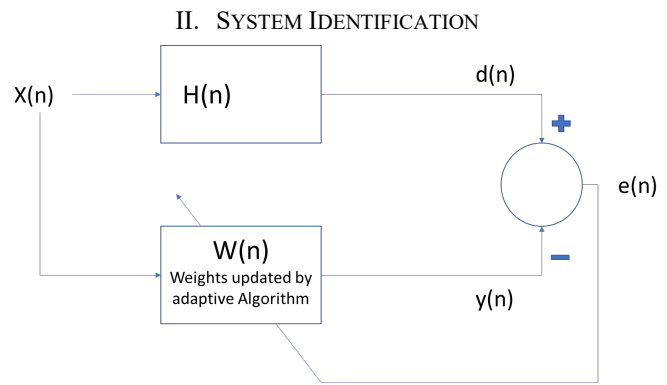

**Figure 1 Block Diagram of System Identification**

System identification is a tool to model unknown systems with finite memory as a finite duration impulse response (FIR) filter and thereby ascertain their filter coefficients. This type of modelling is also known as black box modelling [5]. Initially the input data is sent to the unknown system and the desired output data is collected. The same input data is also sent to the adaptive filter whose coefficients are initially set to zero. The desired and adaptive filter output signals are compared and the error obtained is sent to the adaptive algorithm. The adaptive algorithm updates the filter coefficients to bring it closer to the desired filter. The process is iterative and is repeated until the error is below a acceptable level. At this stage the adaptive filter has "learnt" the equivalent FIR filter model of the unknown system. Further any changes in the input or system is again captured by the adaptive algorithm and the filter coefficients are updated accordingly

## III. LMS Algorithm

The LMS algorithm is an approximation of the popular steepest descent algorithm to achieve faster convergence. Its convergence speed, however, is highly dependent on the eigenvalue spread (conditioning number) of the input-signal autocorrelation matrix. [6].The computational complexity is roughly 2L.

As mentioned in the previous section the signal x[n] is same for both desired filter H and adaptive filter W. The input vector tapped is

$$Shx[n] = (x[n], x[n-1], \ldots\ldots x[n-L+1])$$

Where, L is the length of the adaptive filter.
The basic equations that govern the LMS algorithm are

Output : $y(n) = w^H Shx(n)$
Error : $e(n) = d(n) - y(n)$
Weights : $w(t+1) = w(t) + \mu e^T(n) Shx(n)$

Where, μ is the step size of the algorithm and is decided such that convergence is achieved $0 < \mu < \dfrac{2}{\|x(n)\|^2}$

The computational complexity of the LMS is roughly 2L.

## IV. BNDR-LMS Algorithm

The binormalized data-reusing LMS (BNDR-LMS) algorithm [4] analyzed in this paper uses normalization on two orthogonal directions obtained from current and previous data. Apart from the row input vector

$$Shx[n] = (x[n], x[n-1], \ldots\ldots x[n-L+1])$$

There is also the 2-D input tap vector:

$$Shx2 = \begin{pmatrix} Shx2[n] \\ Shx2[n-1] \end{pmatrix}$$

The basic equations that govern the BNDR-LMS algorithm are

Output : $y(n) = w^H Shx2(1)$
Error : $e(n) = d(n) - y(n)$
Epsilon : $\in = [Shx2^T * Shx2]^{-1} * e(n)$
Weights : $w(t+1) = w(t) + \mu Shx2(1)^* \in$

The step sized used here should be $\mu = \dfrac{1}{\|x(n)\|^2}$ for maximum convergence but sometimes depending on nature of the signal a different step size needs to be used where $0 < \mu < \dfrac{2}{\|x(n)\|^2}$

The computational complexity is roughly $4L + 4I_{inv}$.

Where $I_{inv}$ is the computational complexity of the inverse operation

## V. R-AP Algorithm

The binormalized data-reusing LMS (BNDR-LMS) algorithm [4] analyzed in this paper uses normalization on N orthogonal directions obtained from current and previous data. Apart from this a regularization constant $\partial$ is introduced to prevent singularity of the epsilon matrix

The row input vector of size $1 \times L$ :

$$Shx[n] = (x[n], x[n-1], \ldots\ldots x[n-L+1])$$

There is also input tap vector of size $N \times L$ :

$$ShxN = \begin{pmatrix} ShxN[n] \\ ShxN[n-1] \\ \vdots \\ \vdots \\ ShxN[n-N+1] \end{pmatrix}$$

The basic equations that govern the BNDR-LMS algorithm are

Output : $w(t+1) = w(t) + \mu ShxN(1)^* \in$
Error : $e(n) = d(n) - y(n)$
Epsilon : $\in = [ShxN^T * ShxN + \partial I]^{-1} * e(n)$
Weights : $w(t+1) = w(t) + \mu ShxN(1)^* \in$

The step sized used here should be $\mu = \dfrac{1}{N * \|x(n)\|^2}$ for maximum convergence but sometimes depending on nature of the signal a different step size needs to be used where $0 < \mu < \dfrac{2}{\|x(n)\|^2}$

The value of $\partial$ is very less (close to 0) such that while taking inverse if the matrix is singular the $\partial$ component will make the matrix non- zero.

The computational complexity is roughly $2NL + I_{inv} N^2$

Where $I_{inv}$ is the computational complexity of the inverse operation

## VI. Simulation Results

The unified model presented in [7] has been used to make a comparison of the different algorithms of the AP family.

### A. White Noise

Input signal was taken to be white noise with $\mu = 0, \sigma = 1$.

The desired filter was modelled to be a 13 tap high pass filter with normalized frequency, $f_n = 0.4$. All three algorithms were simulated and the frequency response and mean square performance were analyzed.

$$x(n) = randn(1,T)$$

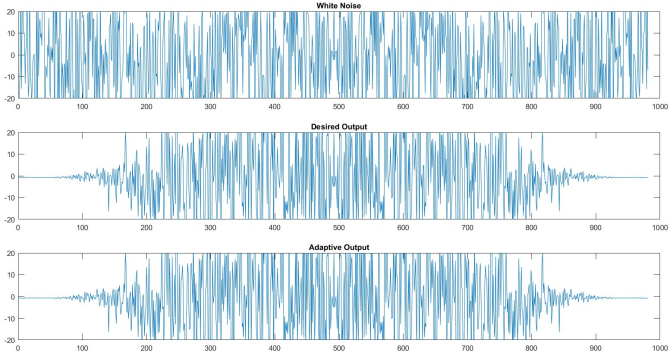

Figure 2 Frequency response of adaptive filter (LMS)

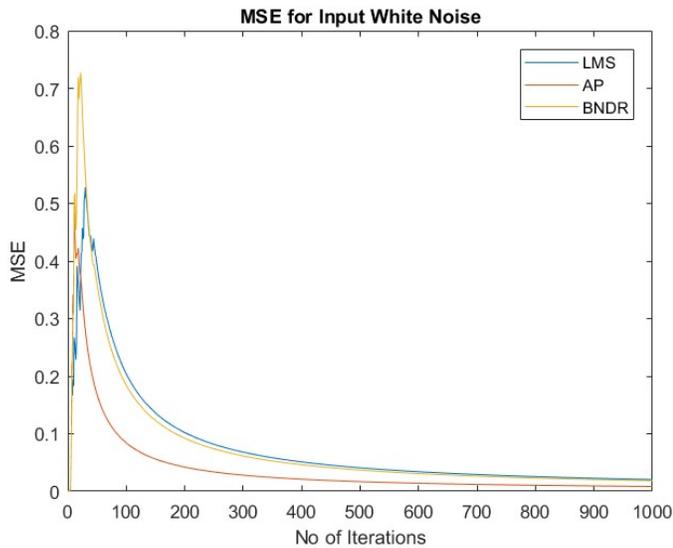

Figure 3 Mean Square Error for white noise

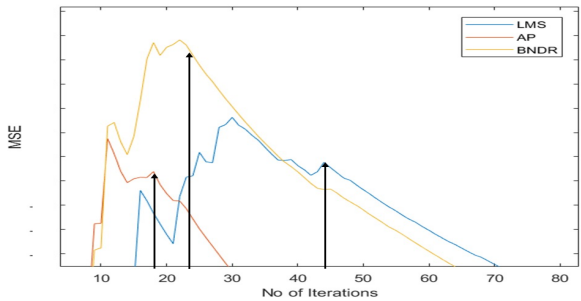

Figure 4 Tm for different algorithms

Let Tm denote the no of iterations the algorithm takes for the MSE to start monotonically decreasing

As seen in Fig 3. the three adaptive algorithms converge quickly and provide the desired frequency response (Fig 2.). The R-AP algorithm converges the fastest and provides the least mean square error. Although the LMS and BNDR-LMS algorithm have similar convergence rates and MSE we can see from Fig 4. that the BNDR-LMS algorithm adapts faster as compared to LMS. It takes the LMS algorithm almost 45 iterations to figure out the trend of the algorithm but the BNDR-LMS algorithm's MSE starts monotonically decreasing by the 20[th] iteration itself.

### B. Coloured Noise

Gray noise of the following frequency response is sent as input to the model:

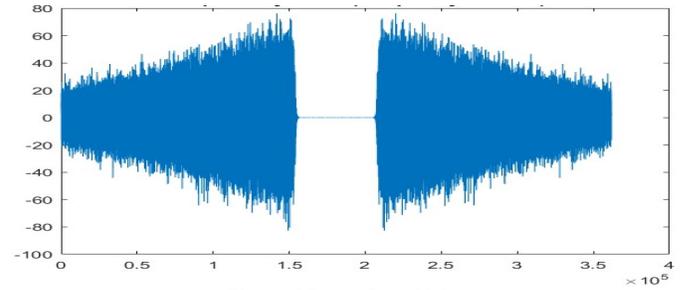

Figure 5 Input Grey Noise

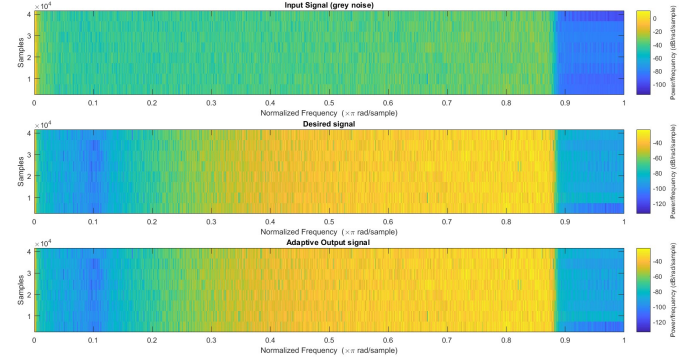

Figure 6 Spectrogram Analysis (R-AP Algorithm)

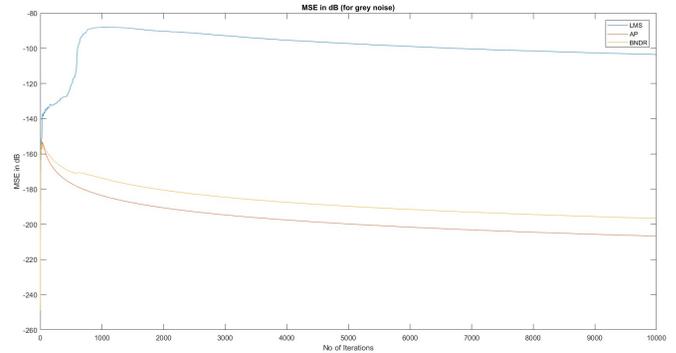

Figure 7 Mean Square Error(MSE) for grey noise

As seen in Fig 7. the LMS algorithm struggles to converge and has a very high MSE(-100dB) as compared to the algorithms in the AP family (-180dB ~ -190dB). Thus, we can see the significant advantage in using the AP family of algorithms as the preferred adaptive algorithm when dealing with applications encountering coloured noise. Fig 6. shows the spectrogram analysis for the R-AP algorithm to validate the fact that system identification is achieved. Moreover, the adaptive filter weights match the desired filter weights as seen in Fig 8.

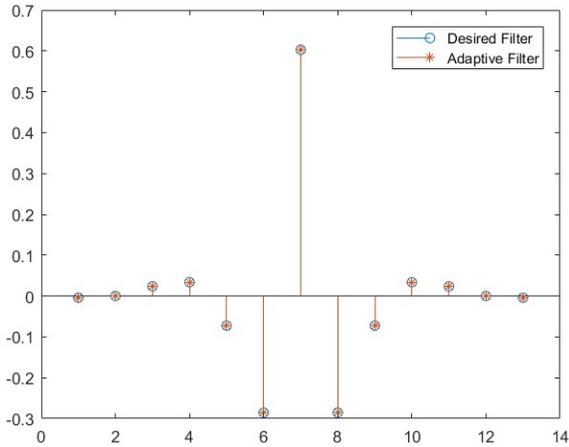

**Figure 8 Adaptive and Desired filter weights**

## VII. Conclusion

In this paper the mean square performance of the conventional LMS and the BNDR-LMS, R-AP Algorithms of the AP family were discussed and verified using simulations. All three algorithms provide low MSE and fast convergence rates for white noise as its input. The R-AP performs the best followed by BNDR-LMS and LMS respectively in regard to mean square performance. But because of the computational complexity and high cost of implementation of the AP family of algorithms, LMS is the preferred algorithm for applications involving white noise. The mean square performance of LMS when encountering grey noise degrades greatly and we see much better performance in the AP family of algorithms. Thus, the R-AP algorithm will be preferred algorithm in applications involving grey noise.

ACKNOWLEDGMENTS

I thank Dr Asutosh Kar (*Asst. Prof.*), Department of Electrical and Electronics Engineering for providing me with this opportunity to work on this topic and providing me valuable advice and suggestions. I also thank Aniruddha Paturkar for clearing my doubts and helping me with any problems encountered during simulations.